\documentclass[%
aps,
prb,
reprint,
superscriptaddress,
amsmath,amssymb,
prresearch,
]{revtex4-2}

\usepackage{siunitx}
\usepackage{amsmath}
\usepackage[utf8]{inputenc}
\usepackage[T1]{fontenc}
\usepackage{graphicx}
\usepackage{lipsum}
\usepackage{layouts}
\usepackage{hyperref}

\begin{document}

\title{Strong nonlocal tuning of the current-phase relation of a quantum dot based Andreev molecule}

\begin{abstract}
	Recent realization of minimal Kitaev chains brought a breakthrough in Majorana research,
	which made arrays of quantum dots coupled by superconductor spacers the most promising synthetic quantum material for topological quantum architectures.
	In this work we investigate the basic building block of this platform --
	two dots coupled via a superconductor (referred to as an Andreev molecule) --
	in a new configuration,
	where two superconducting loops are created to tune the superconducting phase difference across the dots.
	This enables us to take into consideration Coulomb interactions which was not possible in previously studied systems.
	We demonstrate, that the Andreev molecule shows a strong nonlocal Josephson effect:
	as the dot in one junction is tuned the current-phase  relation of the other dot is modified.
	This architecture hosts $0-\pi$ transitions, and shows a tunable anomalous $\varphi_0$ phase-shift,
	nonlocally controlled in both cases, without relying on spin-orbit interaction or Zeeman fields used in previous studies.
	In addition significant superconducting diode effect, and $\pi$-periodic current-phase relations can also be observed.
	The presented nonlocal current-phase relation can be used as a signature of the formation of an Andreev molecular state,
	and in general to introduce new ways to tune quantum architectures.
\end{abstract}

\author{Mátyás Kocsis}
\affiliation{Department of Physics, Institute of Physics, Budapest University of Technology and Economics, Műegyetem rkp. 3., H-1111 Budapest, Hungary}
\affiliation{MTA-BME Superconducting Nanoelectronics Momentum Research Group, Műegyetem rkp. 3., H-1111 Budapest, Hungary}
\author{Zoltán Scherübl}
\affiliation{Department of Physics, Institute of Physics, Budapest University of Technology and Economics, Műegyetem rkp. 3., H-1111 Budapest, Hungary}
\affiliation{MTA-BME Superconducting Nanoelectronics Momentum Research Group, Műegyetem rkp. 3., H-1111 Budapest, Hungary}
\author{Gergő Fülöp}
\affiliation{Department of Physics, Institute of Physics, Budapest University of Technology and Economics, Műegyetem rkp. 3., H-1111 Budapest, Hungary}
\affiliation{MTA-BME Superconducting Nanoelectronics Momentum Research Group, Műegyetem rkp. 3., H-1111 Budapest, Hungary}
\author{Péter Makk}
\affiliation{Department of Physics, Institute of Physics, Budapest University of Technology and Economics, Műegyetem rkp. 3., H-1111 Budapest, Hungary}
\affiliation{MTA-BME Correlated van der Waals Structures Momentum Research Group, Műegyetem rkp. 3., H-1111 Budapest, Hungary}
\author{Szabolcs Csonka}
\affiliation{Department of Physics, Institute of Physics, Budapest University of Technology and Economics, Műegyetem rkp. 3., H-1111 Budapest, Hungary}
\affiliation{MTA-BME Superconducting Nanoelectronics Momentum Research Group, Műegyetem rkp. 3., H-1111 Budapest, Hungary}

\maketitle

\section{Introduction}

Hybrid superconductor-semiconductor structures are the subject of surging fascination,
since they can serve as synthetic quantum materials hosting non-Abelian excitations
\cite{sau_generic_2010, lutchyn_majorana_2010, lutchyn_majorana_2018,beenakker_search_2020, prada_andreev_2020},
and provide topological protection in quantum computational applications
\cite{nayak_non-abelian_2008, flensberg_engineered_2021}.
One of the most promising synthetic quantum material is the Kitaev chain
\cite{sau_realizing_2012}, 
shown in Fig.~\ref{fig:system}a,
consisting of a chain of quantum dots (QD) coupled by superconducting (SC) spacers.
The smallest version of such a chain is two QDs connected to an SC link,
a minimal Kitaev chain hosting states referred to as "poor man's Majorana" states
\cite{leijnse_parity_2012, tsintzis_creating_2022, dvir_realization_2022}.
A similar minimal setup is also used for splitting Cooper-pairs
\cite{hofstetter_cooper_2009,herrmann_carbon_2010,hofstetter_finite-bias_2011,das_high-efficiency_2012,schindele_near-unity_2012,lambert_experimental_2014,tan_cooper_2015},
where the SC-QD coupling is usually weak.
However, when a QD is coupled to an SC, so-called Andreev bound states (ABS) form,
which have been widely studied
\cite{buitelaar_quantum_2002, eichler_even-odd_2007, sand-jespersen_kondo-enhanced_2007, deacon_tunneling_2010, pillet_andreev_2010, maurand_first-order_2012, lee_zero-bias_2012, kumar_temperature_2014, lee_spin-resolved_2014, schindele_non-local_2014, jellinggaard_tuning_2016, gramich_andreev_2017, bretheau_tunnelling_2017, van_woerkom_microwave_2017, hays_direct_2018, laroche_observation_2019, estrada_saldana_temperature_2020}.
When two sites hosting such ABSs are closely spaced, the ABSs hybridize and form an Andreev molecular state,
as described in weak links\cite{pillet_nonlocal_2019,pillet_scattering_2020},
and coupled QDs\cite{scherubl_transport_2019},
and even in multiterminal superconducting devices\cite{kornich_fine_2019,kornich_overlapping_2020,coraiola_hybridisation_2023}.
Recent advancements of nanofabrication allowed the demonstration of the first signatures of such Andreev molecules,
\cite{kurtossy_andreev_2021,coraiola_hybridisation_2023,matsuo_phase-dependent_2023}
and very recently the first observation of poor man's Majorana modes have also been shown.\cite{dvir_realization_2022}

In this work we study the minimal Kitaev chain coupled to two outer SC leads,
as shown by the red dashed rectangle in Fig.~\ref{fig:system}a.
This configuration allows the application of phase biases ($\varphi_L,\varphi_R$) on the two quantum dots,
as well as modifying the level position ($\varepsilon_L,\varepsilon_R$) of the dots (see Fig.~\ref{fig:system}b),
enabling us to examine the role of Coulomb interactions in an Andreev molecule for the first time.
We will show that the presence of the Andreev molecular state induces strong nonlocal current-phase relationship on the dots.

Specifically, we study a device shown in Fig.~\ref{fig:system}b,
two QDs (black) embedded in one SC loop (blue) each,
where the two loops share a side.
Two flux lines (green) can be used to apply arbitrary magnetic flux into the SC loops,
to control the superconducting phase differences across the QDs ($\varphi_L,\varphi_R$).
Adding two side gates (orange) allows us to electrostatically
control the on-site energy of the two QDs separately ($\varepsilon_L,\varepsilon_R$).
This control is not possible if the JJs behave as non-interacting transport channels\cite{pillet_nonlocal_2019,pillet_scattering_2020}.
As we will show this distinction leads to novel behavior in our system.

In the following sections we show how this device behaves in different parameter regimes,
and what robust signatures of the Andreev molecular states can be observed.
The Andreev molecular states are observed through the presence of the nonlocal Josephson effect,
where the supercurrent flowing through one dot is influenced by tuning the parameters of the other QD.
We demonstrate $0-\pi$ and large $\varphi_0$ phase-shifts even in the absence of ground state (GS) change.
$\pi$-periodic CPRs are also demonstrated, for certain parameter configurations.
Unlike previous systems \cite{reynoso_anomalous_2008, buzdin_direct_2008, zazunov_anomalous_2009, goldobin_josephson_2011, sickinger_experimental_2012, yokoyama_anomalous_2014, golubov_current-phase_2004, shukrinov_anomalous_2022, szombati_josephson_2016, mayer_gate_2020},
ours does not rely on spin-orbit interaction (SOI) or a Zeeman field.
This makes such devices especially suited for applications where $\varphi_0$ junctions have been considered,
such as phase batteries\cite{pal_quantized_2019,strambini_josephson_2020}.
We also demonstrate a considerable nonlocally tunable superconducting diode effect.

\section{Methods}
\label{sec:methods}

\begin{figure}[]
	\includegraphics{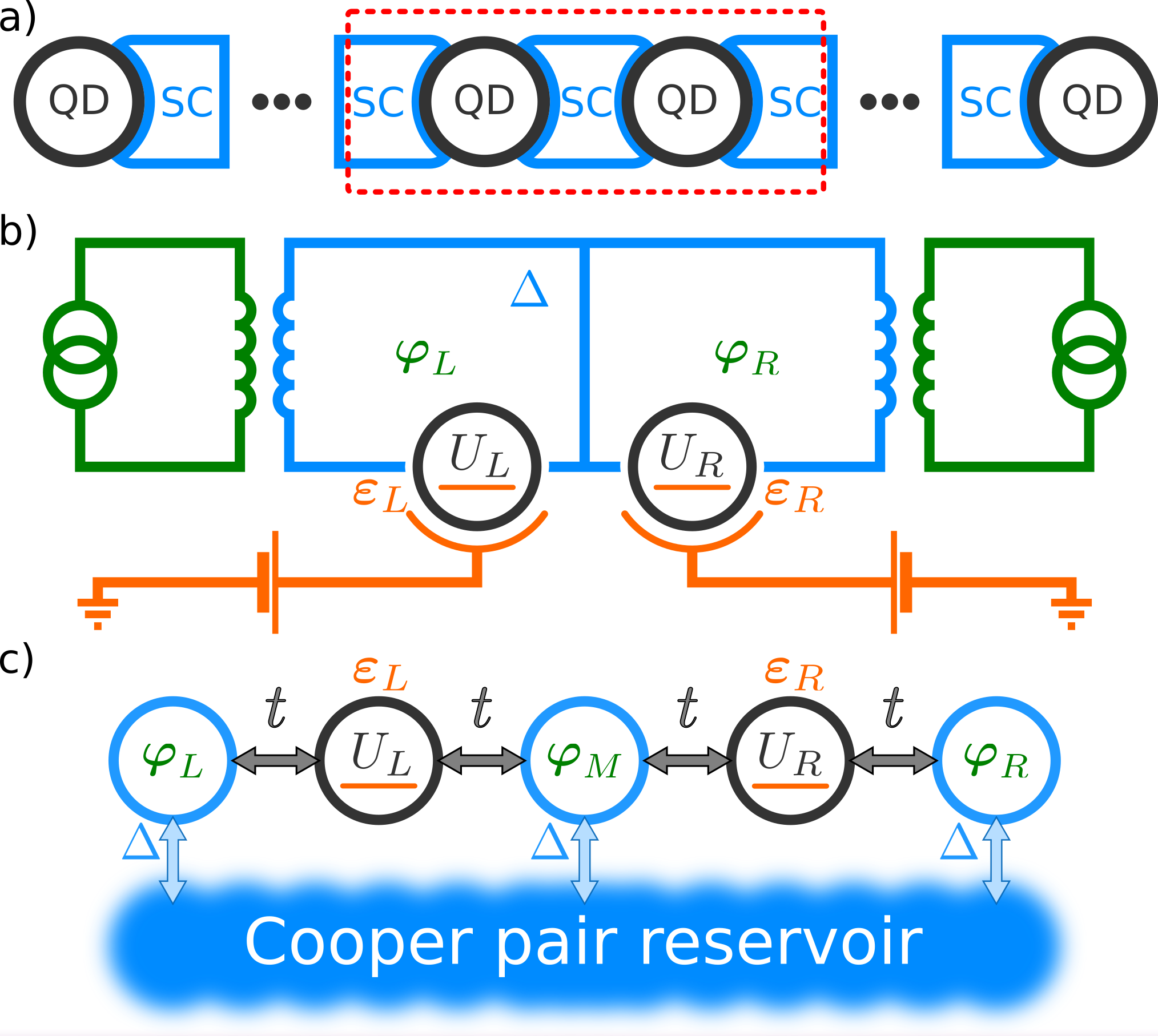}
	\caption[Caption]{
		\emph{a)} A chain of quantum dots (QD) connected by superconductors can host Majorana fermions.
		Three SC sites connected by QDs (dashed red line) can be thought of as a basic building block of such a system.
		\emph{b)} The proposed circuit for studying Andreev molecules, two SQUIDs with a QD in each Josephson junction,
		which allow phase biasing ($\varphi_L,\varphi_R$) across either dot.
		By gating the QDs, and phase-biasing the loops separately we have four individually tunable variables.
		\emph{c)} Five-site model used in our calculations.
		We set the superconducting phase of the middle site to $\varphi_M=0$ in all cases.
		For simplicity, we keep all $t$ hopping terms equal, and set $U_L=U_R=U=1$ as our energy scale, with $\Delta=0.4U$ in all calculations.
	}
	\label{fig:system}
\end{figure}

The phase-biased Andreev molecule system was modeled with a chain of five sites as shown in Fig.~\ref{fig:system}c,
denoted as $\mathrm{SC}_L-\mathrm{QD}_L-\mathrm{SC}_M-\mathrm{QD}_R-\mathrm{SC}_R$,
with all sites coupled to their nearest neighbors.
The left and right QDs are labeled with L and R, while the SC sites are labeled L, M and R for the left, middle and right site respectively.
The coupling strength between nearest neighbors ($t$) is kept the same across the system,
it is scaled with the Coulomb energy of the dots ($U=U_L=U_R=1$),
and so are the on-site energies of the dots ($\varepsilon_R,\varepsilon_L$). 

The Hamiltonian, describing the system can be written as
\begin{equation}
	H = H_{\mathrm{QD}} + H_{\mathrm{SC}} + H_{\mathrm{NN}},
	\label{eq:full}
\end{equation}
where $H_{\mathrm{QD}}$ contains the terms related to the quantum dots,
$H_{\mathrm{SC}}$ the terms related to the superconducting sites,
and $H_{\mathrm{NN}}$ describes the nearest-neighbor tunnel couplings.

We assume that the level spacing of the QDs are large, so each QD can be modeled with a single spinful orbital.
The QDs are treated according to the Anderson model,
\begin{equation}
	H_{\mathrm{QD}}= \sum_{\alpha=\mathrm{QD_L},\mathrm{QD_R}}\varepsilon_\alpha\hat n_{\alpha}+U_\alpha\hat n_{\alpha, \uparrow}\hat n_{\alpha, \downarrow},
\end{equation}
where $\varepsilon_\alpha$ is the on-site energy, and $U_\alpha$ is the on-site Coulomb repulsion energy.
The number operator of the QDs, $\hat n_{\alpha} = \sum_{\sigma=\uparrow,\downarrow}\hat n_{\alpha,\sigma}$,
where $\hat n_{\alpha,\sigma}=\hat c^\dagger_{\alpha,\sigma}\hat c_{\alpha,\sigma}$,
$\hat c_{\alpha,\sigma}$ and $\hat c^\dagger_{\alpha,\sigma}$ are the annihilation and creation operators on site $\sigma$.

When describing the SC leads, we approximate the full BCS Hamiltonian using the zero-bandwidth (ZBW) approximation\cite{steffensen_yu-shiba-rusinov_nodate, gorm_ysr_2021},
where an SC site can only host a single quasiparticle (QP) at energy $\pm\Delta$.
We use the ZBW approximation, as it has been shown to yield results
that compare quantitatively to the outcome of numerical renormalization group calculations (NRG),
when care is taken in choosing the scale of the superconducting gap and the couplings,
describing QDs attached to SCs\cite{steffensen_yu-shiba-rusinov_nodate, gorm_ysr_2021, hermansen_inductive_2022}.
The ZBW Hamiltonian of the SC sites takes the form
\begin{equation}
	H_{\mathrm{ZBW}} = \sum_{\substack{\alpha=\mathrm{SC}_L,\\\mathrm{SC}_M,\mathrm{SC}_R}}\Delta\left(e^{i\varphi_\alpha}\hat c^\dagger_{\alpha, \uparrow} \hat c^\dagger_{\alpha, \downarrow} + \mathrm{h.c.}\right),
	\label{eq:zbw}
\end{equation}
where $\Delta$ is the superconducting gap, $\varphi_\alpha$ is the superconducting phase of the site.
Since the superconducting phase is transferable from one SC site to an other via simple gauge transformations,
we set the superconducting phase on the middle SC site to zero, $\varphi_M=0$.

The nearest-neighbor coupling is expressed as
\begin{equation}
	H_{\mathrm{NN}} = \sum_{\left<\alpha,\beta\right>}t_{\alpha,\beta}\left(\hat c^\dagger_{\alpha, \uparrow}\hat c_{\beta, \uparrow}+\hat c^\dagger_{\alpha,\downarrow}\hat c_{\beta,\downarrow}+\mathrm{h.c.}\right),
\end{equation}
where $t_{\alpha,\beta}$ describes the strength of the coupling between neighboring sites.
For simplicity, we use $t_{\left<\alpha,\beta\right>}=t$ for all $\alpha$ and $\beta$.
Without loss of generality we assume $t$ is real, since the system lacks any SOI. 
Asymmetries in the coupling are discussed in Appendix~\ref{supp:coupling}.
In general, our findings are applicable even if the values of $t_{\alpha,\beta}$ are not precisely matched.

To make sure that the use of the ZBW approximation is valid we always set $t<\Delta<U$.
Unless indicated otherwise, in all calculations $t=0.2U$ and $\Delta=0.4U$.
By using Eq. \eqref{eq:zbw} instead of a full BCS Hamiltonian to describe the SC sites, Eq. \eqref{eq:full} becomes finite dimensional,
so it can be diagonalized numerically.

In this work we consider the ground state (GS) of the system,
which can be divided into even and odd phases, depending on the total number of electrons on the QDs.
This particle parity gives a useful tool,
to explore how the stability diagram of our system is influenced by the strength of the coupling between sites.
It is important to note, that we always consider the whole system,
thus if both QDs have odd occupancy, the system as a whole is still considered to be in an even state.

Our main focus is the supercurrent flowing through the JJs,
which can be calculated by introducing the particle current operators $i\hbar\hat J_\alpha=\partial_t \hat n_\alpha = \left[H,\hat n_\alpha\right]$.
In similar systems with only one QD, the current is calculated using the derivatives of the free energy\cite{gorm_ysr_2021},
but in our case the two separate JJs and three separate currents $J_L, J_M, J_R$ flowing from the SC sites into the the Cooper pair reservoir,
as shown in Fig.~\ref{fig:system}c by the vertical arrows, require the use of the current operators $\hat J_\alpha$.
This enables us to study the three currents separately, as well as the current-phase relations (CPRs) of our system,
and show signatures of the Andreev molecular state.

\section{Results}

\begin{figure}[!h]
	\includegraphics[width=\columnwidth]{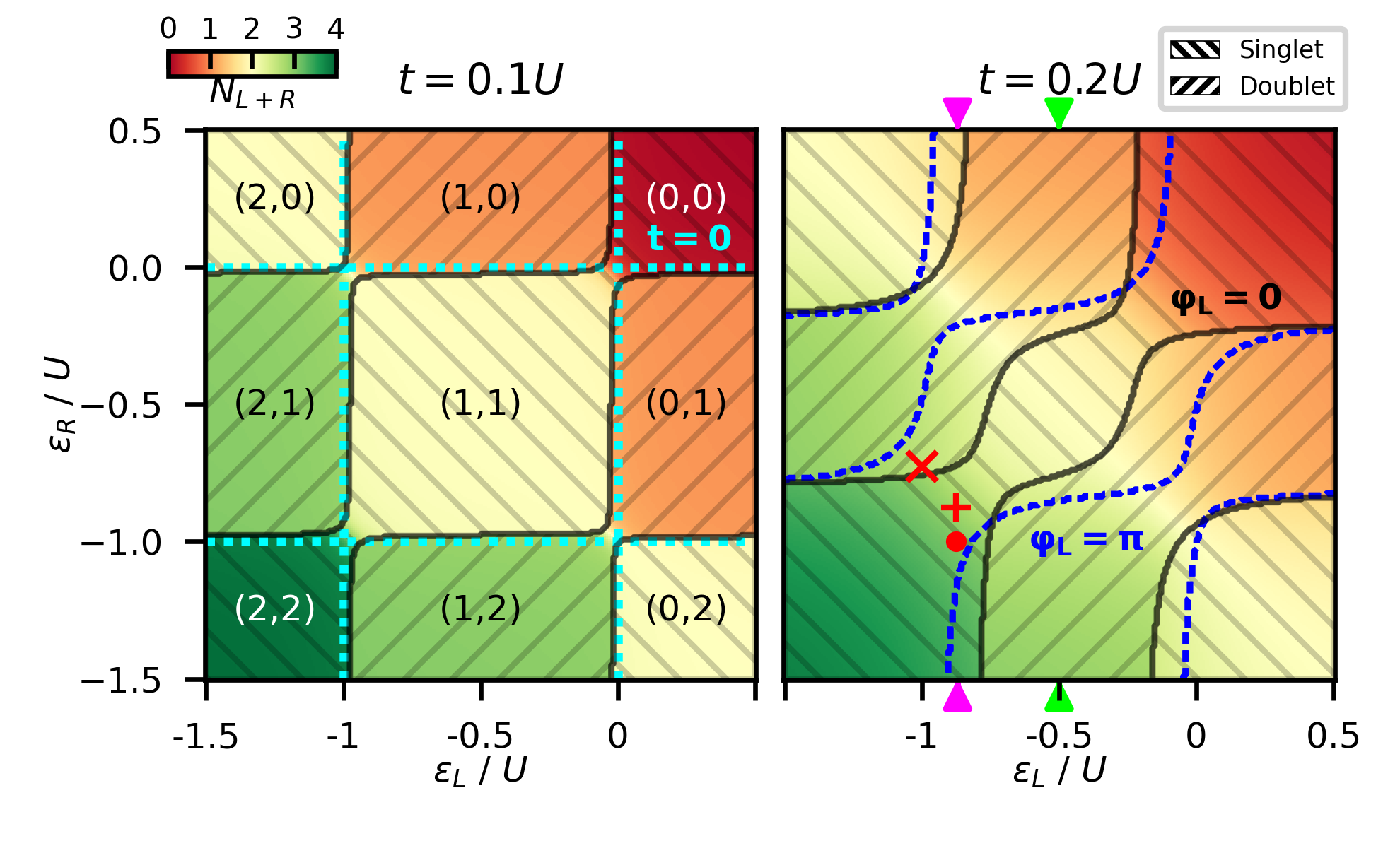}
	\caption[Caption]{
		Stability diagrams at different coupling strengths.
		The colors show the overall particle number as a function of the on-site energy of the two QDs, $\varepsilon_L, \varepsilon_R$.
		The parity of the ground state is indicated by hatching, the solid black line denotes the singlet-doublet boundary.
		\emph{a)}
			Weaker coupling, $t=0.1U$.
			The cyan dotted line shows the singlet-doublet boundary without any coupling, labels show the occupation numbers of left and right QD.
			With nonzero coupling the regions with the same parity hybridize.
			The former (1,1) region shrinks due to the coupling.
		\emph{b)}
			Stronger coupling, $t=0.2U$.
			The hybridized regions expand, the region where (1,1) is the dominant particle number configuration shrinks further.
			The blue dotted line shows the singlet-doublet boundary when a magnetic flux is inserted into the left loop, $\varphi_L=\pi$. 
			In all other cases, no magnetic flux is present, $\varphi_L=\varphi_R=0$.
	}
	\label{fig:eemap}
\end{figure}

\begin{figure}[!h]
	\includegraphics[width=\columnwidth]{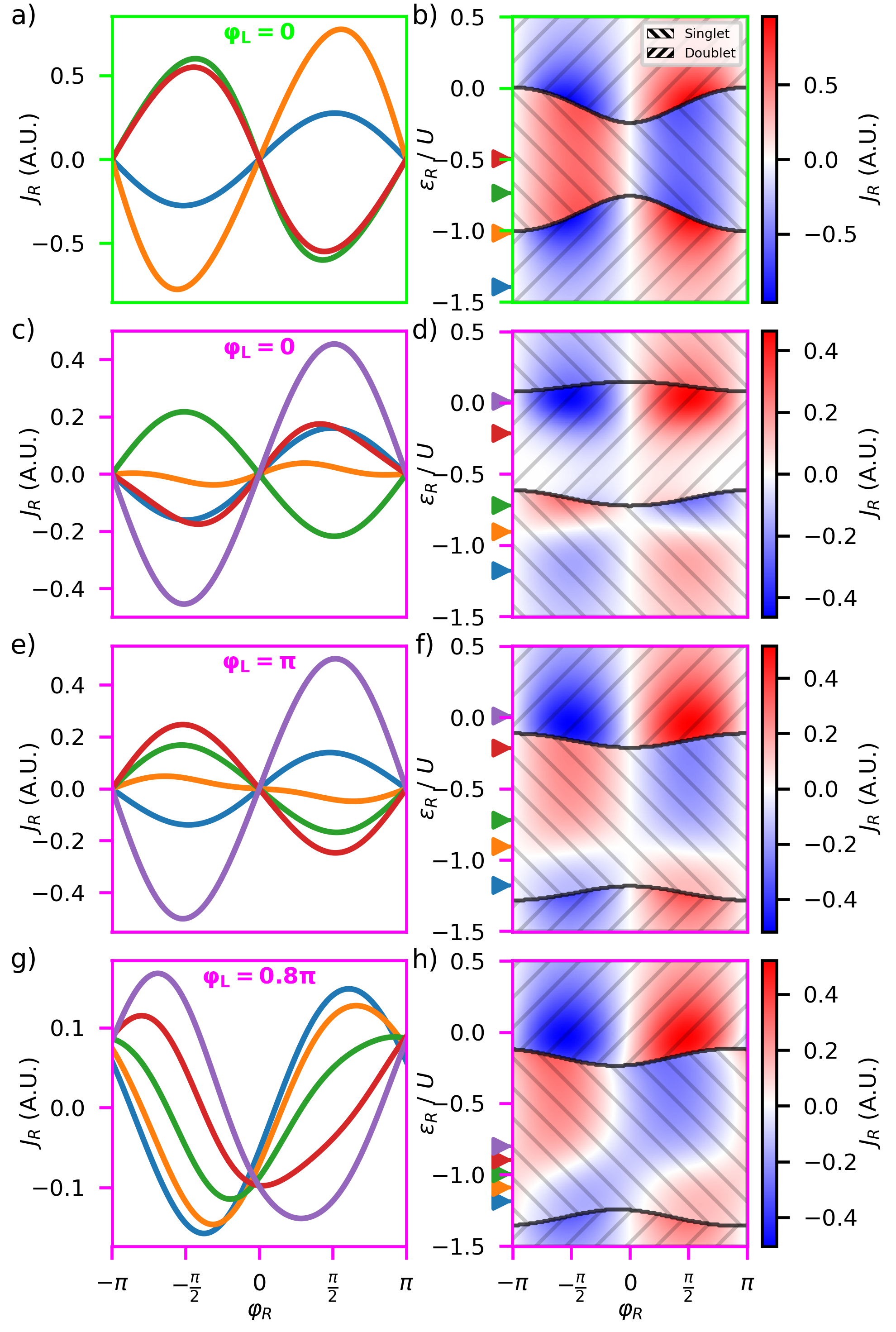}
	\caption[Caption]{
		Current-phase relation (CPR) of the local QD.
		The left column shows curves at some given local on-site energy $\varepsilon_R$,
		the right shows CPRs dependence on the on-site energy.
		Colors represent the local supercurrent $J_R$, hatching indicates the singlet and doublet regions,
		solid lines indicate the phase boundary.
		Arrows indicate the values of $\varepsilon_R$ where the curves on the left were taken.
		\emph{a,b)}
			Nonlocal QD is deep in blockade.
			Blue and green curves show a $0-\pi$ transition,
			the parity of the groundstate changes.
		\emph{c,d)}
			Nonlocal QD is on resonance and $\varphi_L=0$.
			Blue and green curves show a $0-\pi$ transition,
			surprisingly without the need for parity change in the GS.
		\emph{e,f)} 
			Nonlocal QD is on resonance, with $\varphi_L=\pi$.
			The blue and green curves show a $0-\pi$ transition, without GS parity change similar to the case above.
			Singlet and doublet phases are inverted, the transition still takes place in the singlet region.
		\emph{g,h)}
			Nonlocal QD is on resonance, its SC phase is a non-integer multiple of $\pi$ ($\varphi_L=0.8\pi$).
			Curves show a $\varphi_0$ phase-shift, the maximum is continuously shifted with local on-site energy $\varepsilon_R$.
			This $\varphi_0$ transition relies only on the nonlocal phase tuning, and does not require spin-orbit interaction.
	}
	\label{fig:cpr}
\end{figure}

\subsection{Phase diagram}

First let us illustrate (see Fig.~\ref{fig:eemap}a.) how the localized Andreev bound states (ABS) residing on separate dots hybridize into a molecular one,
by studying the charge stability diagram of the system with relatively weak $t=0.1U$ coupling.
The two axes correspond to the on-site energy of the left and right QDs,
the colors show the number of particles in the whole system consisting of two QDs, which can range from 0 to 4.
We label the different regions with particle numbers of the left and right QD, ($n_L, n_R$) which, in the absence of couplings (i.e. $t=0$) would be exact.
Region boundaries of the non-interacting case are marked by the dotted cyan lines in Fig. \ref{fig:eemap}a.
With non-zero coupling strength we still have regions where the labeled states are good approximations of the ground state,
however the boundaries are shifted as marked by the solid lines.
The hatching denotes the odd and even particle parity regions, as disscussed in Section \ref{sec:methods}.
For example the top right region corresponds to both dots being empty hence the (0,0) label, and the hatching denoting the even states.
Since we do not apply any Zeeman field and the system has no SOI, the energy levels of the QDs are spin-degenerate.
For example the state (0,$\uparrow$) has the same energy as (0,$\downarrow$).
The even parity GS is always a singlet, while the odd parity GS is a spin degenerate doublet,
so we use the words singlet (doublet) and even (odd) interchangeably to describe the different regions.

In Fig.~\ref{fig:eemap}a. there are small regions around the corners of the (1,1) charge region where we can see the effects of coupling the QDs to the SC sites,
as avoided corssings.
This coupling shrinks the doublet regions,
e.g. the solid borders of the (1,0), (0,1), regions are shifted inwards from the dotted lines, where they meet.
This is also true for the (2,1), and (1,2) regions,
since single occupancy of the QDs becomes less favored due to the presence of SC correlations on the QDs\cite{van_dam_supercurrent_2006, bargerbos_singlet-doublet_2022}. 
Regions with the same parity start hybridizing,
e.g. the (0,0), (1,1) and (2,2) regions of even parity are connected,
while the odd (1,0) state connects with the (2,1) state.
This is the consequence of crossed Andreev reflection (CAR),
where a Cooper pair from the middle SC site is split up,
and one electron enters the left and right QDs each.
This process couples the localized ABSs residing on the separate dots, to form the molecular Andreev states.

As the strength of the coupling increases,
the doublet regions shrink further, while the hybridized regions expand,
as shown in Fig.~\ref{fig:eemap}b for $t=0.2U$.
Comparing the solid lines on the lower part ($\varepsilon_R<U$) of the two panels of Fig.~\ref{fig:eemap},
we see that the doublet region contracted from spanning the middle half of the axis ($\varepsilon_L\approx-1U$ to $\varepsilon\approx0$) to less than a third,
while the hybridized region of the former (1,2) and (0,1) states expanded significantly.

So far no superconducting phase difference was present across the junctions ($\varphi_L=\varphi_R=0$),
however our systems allows for individually phase biasing each junction.
To demonstrate the effect of flux biasing one of the JJs,
we show the singlet-doublet boundary for the $\varphi_L=\pi,\,\varphi_R=0$ case with a dashed blue line in Fig.~\ref{fig:eemap}b, where the changes are the most pronounced.
Examining the lower part of Fig.~\ref{fig:eemap}b again,
we see that the doublet region has expanded along $\varepsilon_L$,
as $\varphi_L = 0 \rightarrow \pi$
In contrast, the vertical extension of the (0,1) doublet region at $\varepsilon_L=0.5U$ (right side of the panel) is not affected.
Thus remarkably in some regions (e.g. at the point marked with a red $\boldsymbol\times$) a quantum phase transition can be induced,
i.e. a GS parity change can be induced by tuning the superconducting phase.
Similar quantum phase transitions, tuned by the superconducting phase, have been recently observed in simpler \mbox{SC-QD-SC} systems.\cite{bargerbos_singlet-doublet_2022}.

While the charge stability diagrams are useful for understanding the behavior of the system,
experimentally detecting nonlocal effects is most straightforward by measuring the supercurrent flowing through one of the QDs.
In the following we will focus on the current-phase relation (CPR) of the right QD, and show experimental signatures of the molecular state.
We will refer to $\varphi_R$ and $\varepsilon_R$ parameters as local while referring to $\varphi_L$ and $\varepsilon_L$ as nonlocal.
The same effects could also be observed with the roles reversed.
For detailed comparisons of $J_L$ and $J_R$ see Appendix \ref{supp:jl}.

First we will discuss the case when the nonlocal QD is off-resonant,
and our device resembles the simpler \mbox{SC-QD-SC} devices\cite{bargerbos_singlet-doublet_2022}.
This yields a $0-\pi$ transition when the GS changes parity, very similar to the one in the \mbox{SC-QD-SC} setup.
We then move on by tuning the nonlocal QD to resonance, and showing a $0-\pi$ phase-shift of the CPR, \textit{in the absence of} GS parity change.
Then we present how this $0-\pi$ phase-shift appears whenever the SC phase across the nonlocal junction is an integer multiple of $\pi$ ($\varphi_L=k\pi$).
Along the way, we will find that the system can be tuned such that, $\pi$-periodic CPRs can be observed.
Finally, the case of arbitrary $\varphi_L$ is also considered, where we find a tunable $\varphi_0$ phase-shift.

\subsection{Current-phase relations}

\subsubsection{Off-resonance case}

The left QD is tuned far from hybridization between the QDs by tuning $\varepsilon_L$ so that the left QD is deep in blockade.
The green arrows in Fig.~\ref{fig:eemap}b show one such value for $\varepsilon_L$,
equidistant from both resonances at $\varepsilon_L=-0.5U$.
The current-phase relation (CPR) for some values of $\varepsilon_R$ is shown in Fig.~\ref{fig:cpr}a,
while Fig.~\ref{fig:cpr}b shows the same for a wide region of $\varepsilon_R$.
The particular values of $\varepsilon_R$ where the line cuts of Fig.~\ref{fig:cpr}a are taken are indicated by arrows.

The orange and blue curves of Fig.~\ref{fig:cpr}a show a near sinusoidal CPR corresponding to a conventional $0$ junction
(skewness, and higher harmonic components of the CPRs are addressed in Appendix~\ref{supp:harmonics}),
taken at $\varepsilon_R=-1.396U$ and $\varepsilon_R=-1.018U$ respectively.
In this region the GS is a doublet state, as opposed to \mbox{SC-QD-SC} systems where the $0$ junction is in the singlet region\cite{bargerbos_singlet-doublet_2022}.
This is due to the single electron occupying the off-resonance nonlocal QD,
which does not influence the local current, but is counted when determining the particle parity of the whole system,
as discussed in Section~\ref{sec:methods}.

The green ($\varepsilon_R=-0.737$) and red ($\varepsilon_R=-0.496$) curves show CPRs which are shifted by $\pi$ in $\varphi_R$ corresponding to a $\pi$ junction.
The $0-\pi$ transition is driven by the GS transition, yielding a parity change
(as demonstrated by the coincidence of the blue-red color transition and the solid black phase boundary in Fig.~\ref{fig:cpr}b).
This means that we need to add or remove a quasiparticle to/from the system, to observe the $0-\pi$ transition.
As expected these results show the same qualitative behavior as a single \mbox{SC-QD-SC} system\cite{bargerbos_singlet-doublet_2022}, since the left QD is in blockade.

\subsubsection{Hybridization, $\varphi_L=0$}
\label{sss:h1}

Tuning the nonlocal QD towards the hybridization region has very striking effects on the CPRs,
for example by setting $\varepsilon_L=-0.87U$ as shown by the magenta arrow in Fig.~\ref{fig:eemap}b.
Fig.~\ref{fig:cpr}c shows some CPR curves taken at this position.
Comparing the red and green curves we see a $0-\pi$ transition, which is accompanied with a GS parity change as before.
Comparing the blue and green curves we see that a $0-\pi$ phase-shift takes place.
However in strong contrast to the previous example, it is \textit{not} accompanied by a GS parity change.
Fig.~\ref{fig:cpr}d shows the phase-shift between the blue and green arrows taking place entirely in the singlet sector,
which is a direct consequence of the Andreev molecular state spanning the two QDs.
Whenever $0-\pi$ phase-shifts take place without a change in GS parity,
CPR curves with dominant higher harmonic elements can be observed, as shown by the orange curve of Fig.~\ref{fig:cpr}c.
Similar behavior has been predicted and measured in asymmetric $0-\pi$ Josephson junctions consisting of two parallel junctions\cite{goldobin_josephson_2011, sickinger_experimental_2012},
or in balanced SQUID devices \cite{leblanc_nonreciprocal_2023,ciaccia_charge-4e_2023,valentini_parity-conserving_2024}.
In contrast to these devices, which were only tunable by changing the device geometry or by a Zeeman field,
in our case the phase-shift is tuned by local gating.

By choosing all parameters carefully, the CPR can take a close to $\pi$-periodic form in $\varphi_R$,
as shown by the orange curve of Fig.~\ref{fig:cpr}d, and can serve as a unique signature of Andreev molecular states.
Recently proposed protected qubits are based on systems with $\cos 2\varphi$ CPRs\cite{schrade_protected_2022}.
Tuning our system such that the first harmonic part of the CPR is totally supressed,
thus leading to an ideal $\pi$-periodic junction is also possible.
The details of such $\pi$-periodic CPRs, and protected qubits are discussed in Appendix~\ref{supp:harmonics}.

Interestingly, the other GS parity change around $\varepsilon_R\simeq0$ is not accompanied by a $0-\pi$ transition,
rather an $0-0'$ transition transition is taking place, with a significant drop in the amplitude of $J_R$ entering the singlet sector.

\subsubsection{Hybridization, $\varphi_L=\pi$}
\label{sss:h2}

By tuning the nonlocal flux to $\varphi_L=\pi$, we see a dramatic shift of the CPRs from that of Fig.~\ref{fig:cpr}d to Fig.~\ref{fig:cpr}f.
Such dramatic dependence of the local CPR on the nonlocal flux is also a characteristic signature of the Andreev molecular state.
To understand how this change manifests, we come back to the stability diagram.

When comparing the blue dashed line of Fig.~\ref{fig:eemap} corresponding to $\varphi_L=\pi$,
with the solid lines corresponding to $\varphi_L=0$,
we see that the singlet and doublet regions have flipped.
Now the central region is a singlet hybrid of the (2,2) and (1,1) states,
and the outer regions are in doublet GSs.
This inversion of the parity regions is most notable when comparing the hatching of Fig.~\ref{fig:cpr}d and Fig.~\ref{fig:cpr}f. 
The CPR curves in Fig.~\ref{fig:cpr}e show the same $0-\pi$ phase-shift without a GS transition as discussed earlier (see blue and green curves).
The orange curves of Fig.~\ref{fig:cpr}c and Fig.~\ref{fig:cpr}e show similar close to $\pi$-periodic CPRs in the singlet sector.

\subsubsection{$\varphi_0$ phase-shift}

In all cases discussed so far the nonlocal phase was either $0$ or $\pi$,
however tuning $\varphi_L$ to non-integer multiples of $\pi$ can yield exciting new features.
This is demonstrated in Fig.~\ref{fig:cpr}g and h for $\varphi_L=0.8\pi$,
where instead of a $0-\pi$ phase-shift, the phase of the CPRs in the singlet region is shifted by an arbitrary phase $\varphi_0$.

Josephson junctions in which the critical current takes on an anomalous phase, such that $J_c=J_0\sin\left(\varphi+\varphi_0\right)$,
have been studied both theoretically and experimentally\cite{reynoso_anomalous_2008, buzdin_direct_2008, zazunov_anomalous_2009, goldobin_josephson_2011, sickinger_experimental_2012, yokoyama_anomalous_2014, golubov_current-phase_2004, shukrinov_anomalous_2022},
and are a great candidate for the creation of phase batteries\cite{pal_quantized_2019,strambini_josephson_2020}.
In some cases the value of $\varphi_0$ is even tunable\cite{szombati_josephson_2016, mayer_gate_2020, strambini_josephson_2020}.
However, in all cases SOI or a Zeeman field are required.
In our case neither SOI or external fields are required to produce this anomalous phase-shift, and tune the value of $\varphi_0$.
This tunable phase-shift in the absence of SOI or external fields is a strong indicator of the presence of Andreev molecular states.

In systems with a single junction, time-reversal symmetry (TRS) dictates that $J(\varphi)=-J(-\varphi)$,
which implies that $J(\varphi=0)=0$.
The presence of an anomalous Josephson current $J(\varphi=0)\neq0$, can only occur if TRS is broken, for example by SOI or a Zeeman field\cite{shukrinov_anomalous_2022}.

For our system TRS dictates that $J(\varphi_L,\varphi_R)=-J(-\varphi_L,-\varphi_R)$\cite{pillet_nonlocal_2019},
which in the case of $\varphi_L=k\pi$ simplify to $J(\varphi_R)=-J(-\varphi_R)$.
The vertical white bands in the middle of Fig.~\ref{fig:cpr}b,d,f at $\varphi_R=0$ show this symmetry.
If $\varphi_L$ is set to an arbitrary value, the $J(\varphi_R=0)=0$ no longer holds true.
This effect is demonstrated in Fig.~\ref{fig:cpr}g and h for $\varphi_L=0.8\pi$,
where a $\varphi_0$ phase-shift is observed in the singlet region.
The anomalous phase $\varphi_0$ is also strongly tunable by local gating.

Having a non-integer multiple of $\pi$ as the nonlocal phase, also introduces significant changes in the shape of the CPR curves.
The CPR curves shown up to now all had a symmetry where the absolute value of the minimal and maximal supercurrent was equal \mbox{$\left|\mathrm{max}\left(J_R\left(\varphi_R\right)\right)\right|=\left|\mathrm{min}\left(J_R\left(\varphi_R\right)\right)\right|$} (see Fig.~\ref{fig:cpr}a,c,e).
For the curves of Fig.~\ref{fig:cpr}g however 
\mbox{$\left|\mathrm{max}\left(J_R\left(\varphi_R\right)\right)\right|\neq\left|\mathrm{min}\left(J_R\left(\varphi_R\right)\right)\right|$},
with the green curve showing the strongest effect (with the absolute value of the minimum and maximum showing a 28\% difference).
To observe such effect in other systems, both inversion symmetry and TRS has to be broken,
and is referred to as the superconducting diode effect\cite{reynoso_anomalous_2008, reynoso_spin-orbit_2012, wakatsuki_nonreciprocal_2017, mazur_gate-tunable_2022, baumgartner_supercurrent_2022, wu_field-free_2022,gupta_gate-tunable_2023,ciaccia_gate-tunable_2023,baumgartner_supercurrent_2022,pillet_josephson_2023,nadeem_superconducting_2023}.

\subsubsection{Nonlocal phase tuning}

All three effects discussed so far are signatures of the Andreev molecular states formed in the QDs.
Since tuning the nonlocal phase has such a fundamental effect on the system,
we will study the $\varphi_0$ and $0-\pi$ phase-shifts in more detail as a function of the nonlocal phase.
We then show a third scenario, where the nonlocal phase drives a singlet-doublet transition.

\begin{figure}[]
	\includegraphics[width=\columnwidth]{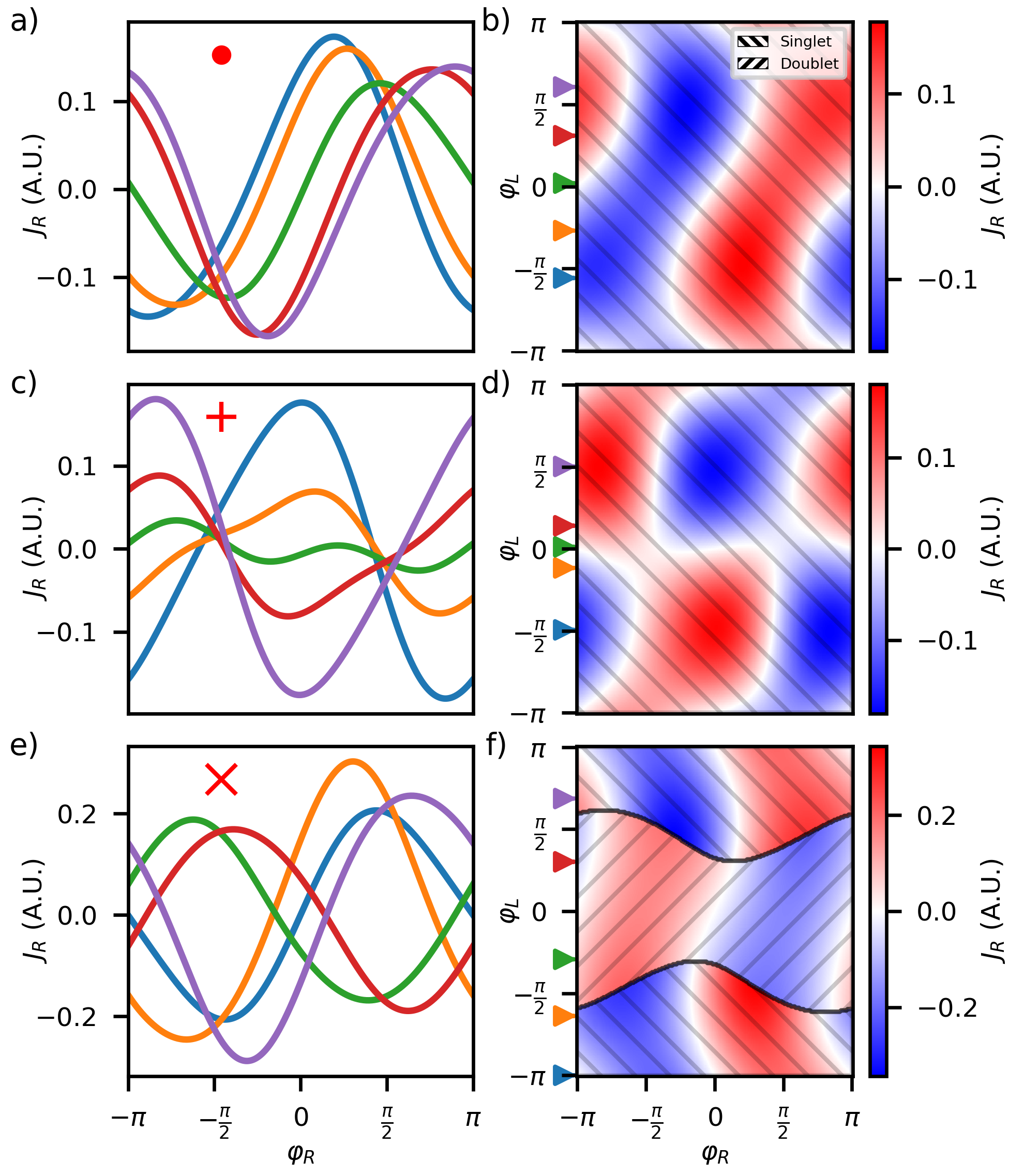}
	\caption[Caption]{
		Current-phase relation (CPR) of the local QD, as a function of the nonlocal superconducting phase $\varphi_L$.
		The nonlocal phase tuning of the local supercurrent is the nonlocal Josephson effect.
		The on-site energies where the CPRs are taken are marked with corresponding red marks in Fig.~\ref{fig:eemap}.
		The left column shows curves at some given nonlocal superconducting phase values $\varphi_L$,
		while the right column shows how CPR depends on the nonlocal superconducting phase in a continuous window.
		\emph{a,b)}
			$\varphi_0$ phase-shift, tuned by nonlocal phase.
			The curves are continuously shifted by the nonlocal phase.
			The absolute value of the minimum and maximum current differs for each curve, showing the superconducting diode effect.
			This is true for all on-site energy configurations shown here.
		\emph{c,d)}
			$0-\pi$ phase-shift, tuned by nonlocal phase, without requiring the change of GS parity.
			Close to the transition the CPR is strongly nonsinusoidal.
		\emph{e,f)}
			The quantum phase transition of the GS from the singlet to the doublet state is driven by the nonlocal phase.
			Within the same parity regions $\varphi_0$ phase-shift is observable,
			while $0-\pi$ transitions take place along the singlet-doublet boundary.
	}
	\label{fig:nljf}
\end{figure}

By fixing the on-site energies to the values indicated in Fig.~\ref{fig:eemap} by a red dot,
we can study how the nonlocal phase $\varphi_L$ tunes the $\varphi_0$ shift of the local CPR.
Fig.~\ref{fig:nljf}a shows a few selected CPR curves at different values of $\varphi_L$, demonstrating the $\varphi_0$ junction like behavior.
The exact phase shift can be tuned in a wide range by the nonlocal phase.
When both local and nonlocal phases are zero the current is completely suppressed.

Mirroring Fig.~\ref{fig:nljf}b around the $\varphi_L=0,\varphi_R=0$ point, and inverting the colors yields the original figure.
This is a consequence of TRS mentioned earlier,
which implies that $J(\varphi_L,\varphi_R)=-J(-\varphi_L,-\varphi_R)$\cite{pillet_nonlocal_2019}.

A $\pi$ phase-shift controlled by the nonlocal phase is also achievable by tuning the dots,
such that $\varepsilon_L=\varepsilon_R$ as shown by the red $\boldsymbol+$ in Fig.~\ref{fig:eemap}.
Fig.~\ref{fig:nljf}c and d show the local current reversal by nonlocal phase.
The nonlocal phase switches the junction from a $0$ to a $\pi$ phase-shift,
as the blue and purple curves show, without changing the parity of the GS.
The change from the $\varphi_0$ to the $0-\pi$ regime is detailed in \ref{supp}.

It is also possible to drive GS transition between the singlet and doublet groundstate by nonlocal flux tuning, as shown in Fig.~\ref{fig:nljf}e and f.
To achieve this we set the on-site energies of the QDs close to the boundary, as shown in Fig.~\ref{fig:eemap} by a red $\boldsymbol\times$.
The GS switch also means a $0-\pi$ transition, while within a given GS the nonlocal phase has a $\varphi_0$ like behavior.
Comparing the red and green curves of Fig.~\ref{fig:nljf}e we see that the SC diode effect has the same strength, but the polarity is flipped.
This means that in this regime the system can be used as an SC diode in which the strength and the polarity of the effect is easily tunable.

\section{Discussion}

Let us now compare our QD-based Andreev molecule with a system with the same geometry,
but where the JJs are modeled as non-interacting channels\cite{pillet_nonlocal_2019, pillet_scattering_2020}.
There are three phenomena discussed in our work that are also present in the non-interacting-channel-based model.
These are the breaking of the $J_R(\varphi_R=0)\neq0$ symmetry, the superconducting diode effect, and the tunable $\varphi_0$ phase-shift.
In addition to being able to control the $\varphi_0$ shift of the transition via the nonlocal phase $\varphi_L$ (Fig.~\ref{fig:nljf}b and f),
our system allows it to be controlled via the local gate voltage (onsite energy) $\varepsilon_R$ (Fig.~\ref{fig:cpr}h) as well,
in stark contrast to the non-interacting case.
It is important to note that our model does not consider the distance between the two junctions,
which is an important parameter of the experimental realization.

The $0-\pi$ phase-shift without changing GS parity, are absent in the non-interacting model, they are unique to our QD-based one.
These are markedly different from $0-\pi$ transitions where the GS parity changes:
there exists a central region between the $0$ and $\pi$ phases,
where the CPR is nonsinusoidal and the amplitude of the supercurrent is low,
as opposed to the sharp changes characteristic of GS-changing transitions.
This holds true whether the transition is tuned by the local on-site energy $\varepsilon_R$ (Fig.~\ref{fig:cpr}d and f)
or the nonlocal phase $\varphi_L$ (Fig.~\ref{fig:nljf}d).

In this paper we have studied a QD-based Andreev moelcule between SC leads.
We explored its characteristics for different level positions of the dots and different phase biasing of the JJs,
which lead to unusual and strongly varying current-phase relationship due to the molcular hybridization.

Our proposed circuit can be fabricated from state-of-the-art semiconductor/superconductor nanostructures,
like nanowires with an epitaxial Al shell, that is etched away to form the QDs,
or 2DEG systems proximitized with an epitaxial Al layer.
Both platforms have been used experimentally to create similar devices\cite{kurtossy_andreev_2021, pita-vidal_direct_2022, matsuo_observation_2022, wesdorp_microwave_2022, lee_transport_2019},
showing the feasibility of the realization of the device concept  under investigation. 
The semiconductor can be depleted by local gating, which allows the characterization of a single JJ at a time.
Local gating can tune the energy level of the QD in the presence of Coulomb interactions, which has been crucial in many recent experiments as well\cite{kurtossy_andreev_2021, pita-vidal_direct_2022, matsuo_observation_2022}.
The supercurrents can be measured using high frequency techniques, commonly used in measuring similar devices\cite{wesdorp_microwave_2022, pita-vidal_direct_2022, bargerbos_spectroscopy_2022, bargerbos_singlet-doublet_2022}

We demonstrated how tuning the nonlocal QD away from Coulomb-blockade results in $0-\pi$ or $\varphi_0$ phase-shift of the CPR of the local JJ.
The nonlocal Josephson effect is demonstrated by showing how the nonlocal flux can influence the behavior of the local current.
This yields $0-\pi$ and $\varphi_0$ phase-shifts as well.
Contrary to the single dot case, the nonlocally controlled $0-\pi$ and $\varphi_0$ phase-shifts can occur without quasiparticle parity changes,
and a significant and highly tunable SC diode effect is also demonstrated.

Unlike devices that show similar behavior, our system does not rely on Zeeman fields, or SOI.
Our results show that these effects can be observed in a wide parameter range.
This makes the system valuable for studing both the superconducting diode effect,
and the applications of a programmable $\varphi_0$ junction.
The strong nonlocal tuning of the current-phase relation is a hallmark of the Andreev molecular state,
which is promising for future quantum architectures, like protected qubits. 

In this work we studied the ground state properties of the system,
studying the excitation spectrum may reveal further experimentally observable features of the Andreev molecular state.

It has come to our atttention, that due to high interest in the field,
during the preparation of this manusript multiple new studies have been carried out\cite{haxell_demonstration_2023,matsuo_engineering_2023,matsuo_josephson_2023,pillet_josephson_2023,hodt_-off_2023}.

\section{Data availability}

All data is available upon request from the author.

\begin{acknowledgements}
	The Authors declare no Competing Financial or Non-Financial Interests.
	This work has received funding from
	SuperGate Fet Open,
	the FET Open AndQC,
	Twistrain ERC,
	and from the OTKA grants,
	FK-132146, K-138433, K-134437.
	This research was supported by the Ministry of Culture and Innovation and the National Research, Development and Innovation Office within the Quantum Information National Laboratory of Hungary (Grant No. 2022-2.1.1-NL-2022-00004),
	Grant Nr. TKP2021-NVA-02,
	by the ÚNKP-22-5, and
	the ÚNKP-23-5-BME-413 New National Excellence Program of the Ministry for Culture and Innovation from the source of the National Research, Development and Innovation Fund,
	the János Bolyai Research Scholarship of the Hungarian Academy of Sciences,
	and the EIC Pathfinder Challenge grant QuKiT (Grant number: 101115315).	
\end{acknowledgements}

\section*{Authors Contribution}
M. K. developed the simulations with the help of G. F. and Z. S..
M. K. ran the simulations and visualizations.
P. M. and Sz. Cs. supervised the work.
All authors analyzed and discussed the results, and contributed to the manuscript.

\appendix
\label{supp}
\section{Asymmetric coupling}
\label{supp:coupling}

\begin{figure*}[]
	\includegraphics[width=2\columnwidth]{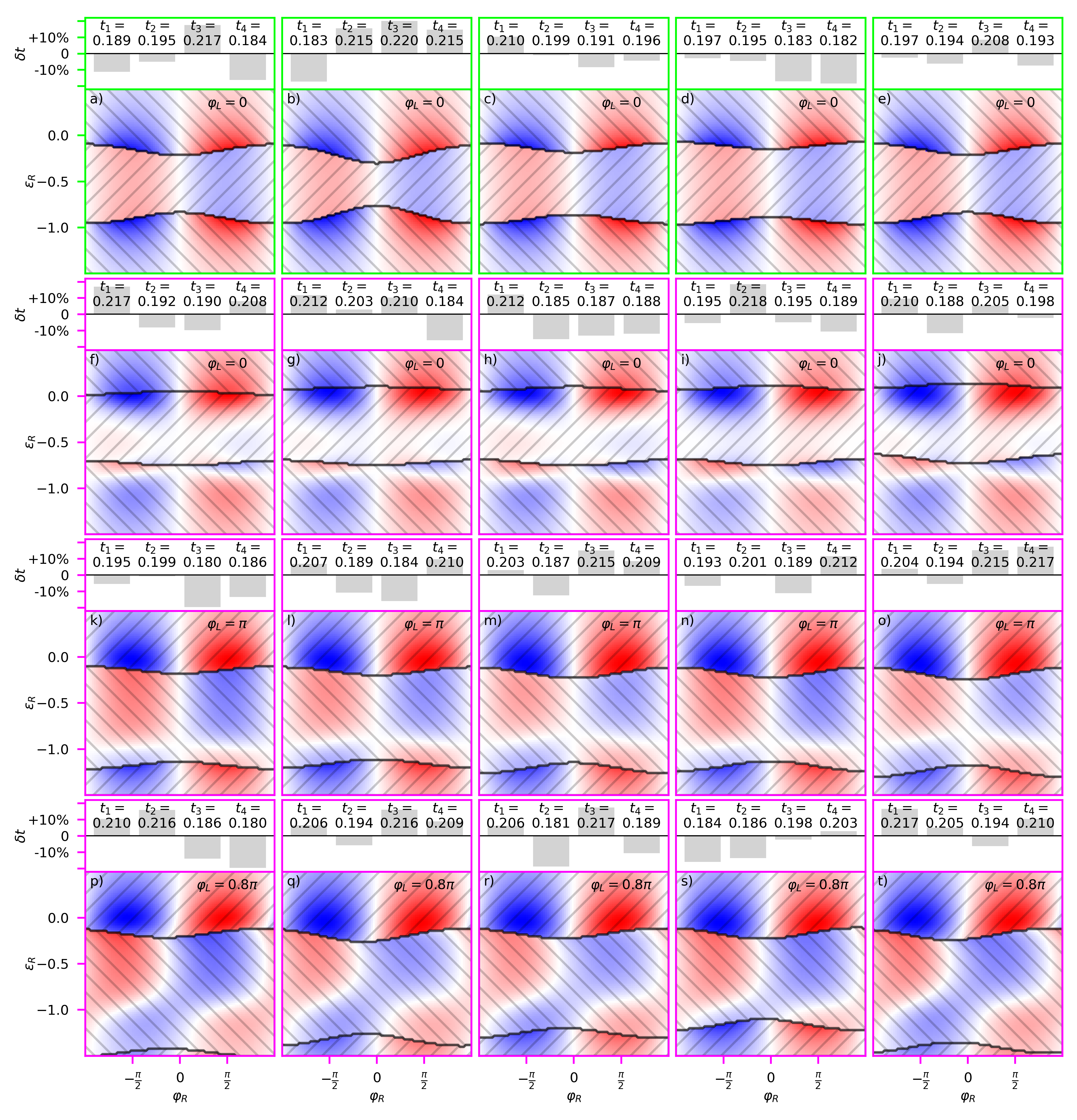}
	\caption[Caption]{
		Local current $J_R$ as a function of the local phase $\varphi_R$ and the local on-site energy $\varepsilon_R$,
		at the same nonlocal phase $\varphi_L$ and nonlocal on-site energy $\varepsilon_L$ values as Fig.~\ref{fig:cpr}.
		For each row represents a parameter setting, with the couplings $t_{\alpha,\beta}$ randomized within a $\pm20\%$ range five times.
		The main features discussed in the main text are conserved even for randomized couplings,
		they are not a consequence of symmetric values of $t_{\alpha,\beta}$.
		The color scales of each panel are normalized individually, the colors of different panels are not comparable.
	}
	\label{figs:t}
\end{figure*}

So far we have assumed whether all coupling strengths are equal, $t_{\alpha,\beta}=t$.
In practice, matching all couplings exactly might not be feasible,
thus it is important to investigate that our findings hold for systems with different couplings.
We reproduced Fig.~\ref{fig:cpr}, with the couplings randomly varied in a window $\pm20\%$ of the original $t=0.2$ value shown in Fig.~\ref{figs:t}.
The coupling terms $t_{\alpha,\beta}$ are numbered left to right as shown by the double sided gray arrows on Fig.~\ref{fig:system}c.
$t_1$ ($t_2$) determines the coupling strength between the left QD and the left SC (middle SC),
while $t_3$ ($t_4$) represent the coupling between the right QD and the middle SC (right SC).
Each row of Fig.~\ref{figs:t} corresponds to a row of Fig.~\ref{fig:cpr}, with the same on-site energy and phase settings,
but each panel of the row is generated with randomized values for $t_{\alpha,\beta}$, the exact values are shown above the panels.
The color scale of each panel is normalized to that single panel, to make all features visible.
This makes the colors of different panels incomparable.

The first row shows the \mbox{SC-QD-SC} like behavior, with two $\pi$ phase shifts when the GS changes parity.
We expect that the role of the superconducting phase between the two SC sites ($\varphi_R$, between the middle and right SC site) will have a stronger effect,
when the coupling between the SC sites and the QD is stronger.
Since the left QD is not on resonance, we only have to consider $t_3$ and $t_4$, the coupling of the right QD to the two neighboring SC sites.
When the couplings are strong, i.e. Fig.~\ref{figs:t}b, we see that the local phase has a strong effect on the width of the doublet region,
it is much narrower at $\varphi_R=0$ than at $\varphi_R=\pi$.
When the couplings are weak, i.e. Fig.~\ref{figs:t}d, the width of the doublet region is much less affected by the phase,
but the two cases are qualitatively the same.

The second row shows the case where the left dot is on resonance, with no phase applied to the nonlocal QD $\varphi_L=0$.
Here we see the $\pi$ phase shift in the doublet region around $\varepsilon_R=-0.8\mathrm{U}$, discussed earlier, appear for all couplings.

The third and fourth row show the cases where a finite phase is applied to the nonlocal QD, $\varphi_L\neq0$.
Here again we see the same features of Fig.~\ref{fig:cpr}, with the exact positions of the features shifting,
but still showing a good qualitative agreement.

\section{Supercurrent of the nonlocal QD}
\label{supp:jl}

\begin{figure}[]
	\includegraphics[width=\columnwidth]{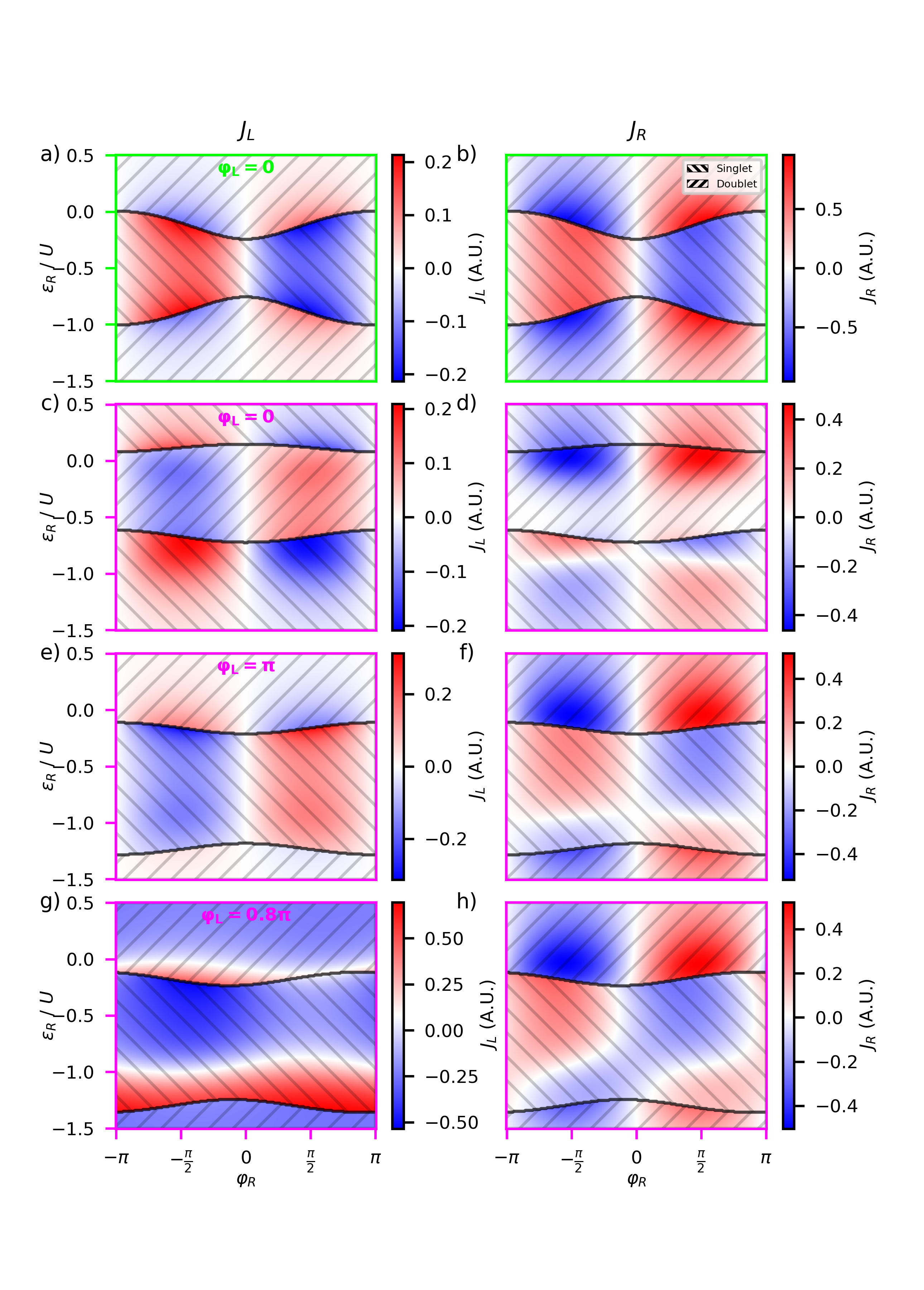}
	\caption[Caption]{
		The current of the nonlocal junction $J_L$, and the local junction $J_R$,
		as a function of the local phase $\varphi_R$ and the local on-site energy $\varepsilon_R$.
		For each row, the nonlocal parameters $\varphi_L$ and $\varepsilon_L$ are the same as for the same row of Fig.~\ref{fig:cpr}.
	}
	\label{figs:jl1}
\end{figure}

\begin{figure}[]
	\includegraphics[width=\columnwidth]{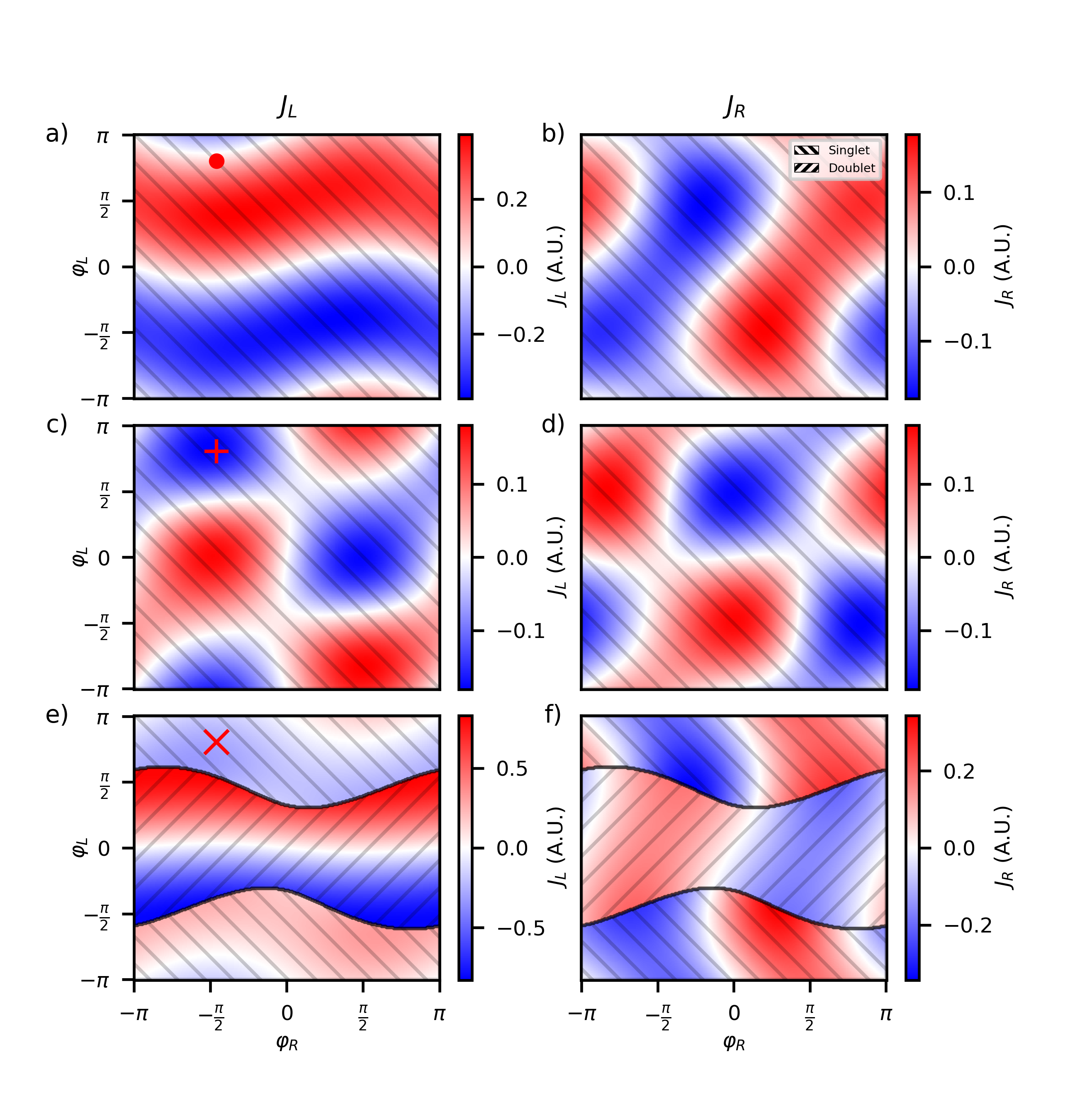}
	\caption[Caption]{
		The current of the nonlocal junction $J_L$, and the local junction $J_R$,
		as a function of the local ($\varphi_R$) and nonlocal ($\varphi_L$) phase.
		For each row, the on-site energies are the same as for the same row of Fig.~\ref{fig:nljf}.
		The on-site energies where the CPRs are taken are marked with corresponding red marks in Fig.~\ref{fig:eemap}.
	}
	\label{figs:jl2}
\end{figure}

So far we only concerned ourselves with the supercurrent flowing through the local QD $J_R$,
however the role of local and nonlocal QD was arbitrarily set,
and the roles could be easily reversed.
In this section we reproduce Fig.~\ref{fig:cpr} and Fig.~\ref{fig:nljf} with both $J_R$ and $J_L$ shown.

Fig.~\ref{figs:jl1}a shows $J_L$, when the left QD is not on resonance.
Comparing the color bars of Fig.~\ref{figs:jl1}a and b, we see that $J_L$ is much smaller than $J_R$,
as expected.
The supercurrent in the left QD is also suppressed, when the SC phase is set to $k\pi$, as shown in Fig.~\ref{figs:jl1}c and e.
If the left QD were not part of a larger system, but the single QD of a \mbox{SC-QD-SC} system,
we would expect the SC current to be zero.
The fact that there is finite current, and its amplitude is tuned by the parameters of the other QD ($\varphi_R$,$\varepsilon_R$),
is further evidence of the Andreev molecular states.

Fig.~\ref{figs:jl2}a and b show that when the on-site energies are tuned to the values shown by the red dot of Fig.~\ref{fig:eemap},
both QDs show the $\varphi_0$ phase-shift discussed in the main text.
When comparing panel a and b of Fig.~\ref{figs:jl2} we have to remember that if we wanted to reverse the roles of the local and nonlocal QDs of panel a,
in essence we would have to exchange the $\varphi_L$ and $\varphi_R$ axes, which would mean mirroring the image along the $\varphi_L=\varphi_R$ diagonal.
This is even more obvious for panels c and d, where $\varepsilon_L=\varepsilon_R$ as shown by the red $\boldsymbol+$ of Fig.~\ref{fig:eemap},
and correspondingly, there is no distinction between the dots, mirroring one panel yields the other.

\section{Higher harmonics of the supercurrent}
\label{supp:harmonics}

\begin{figure*}[]
	\includegraphics{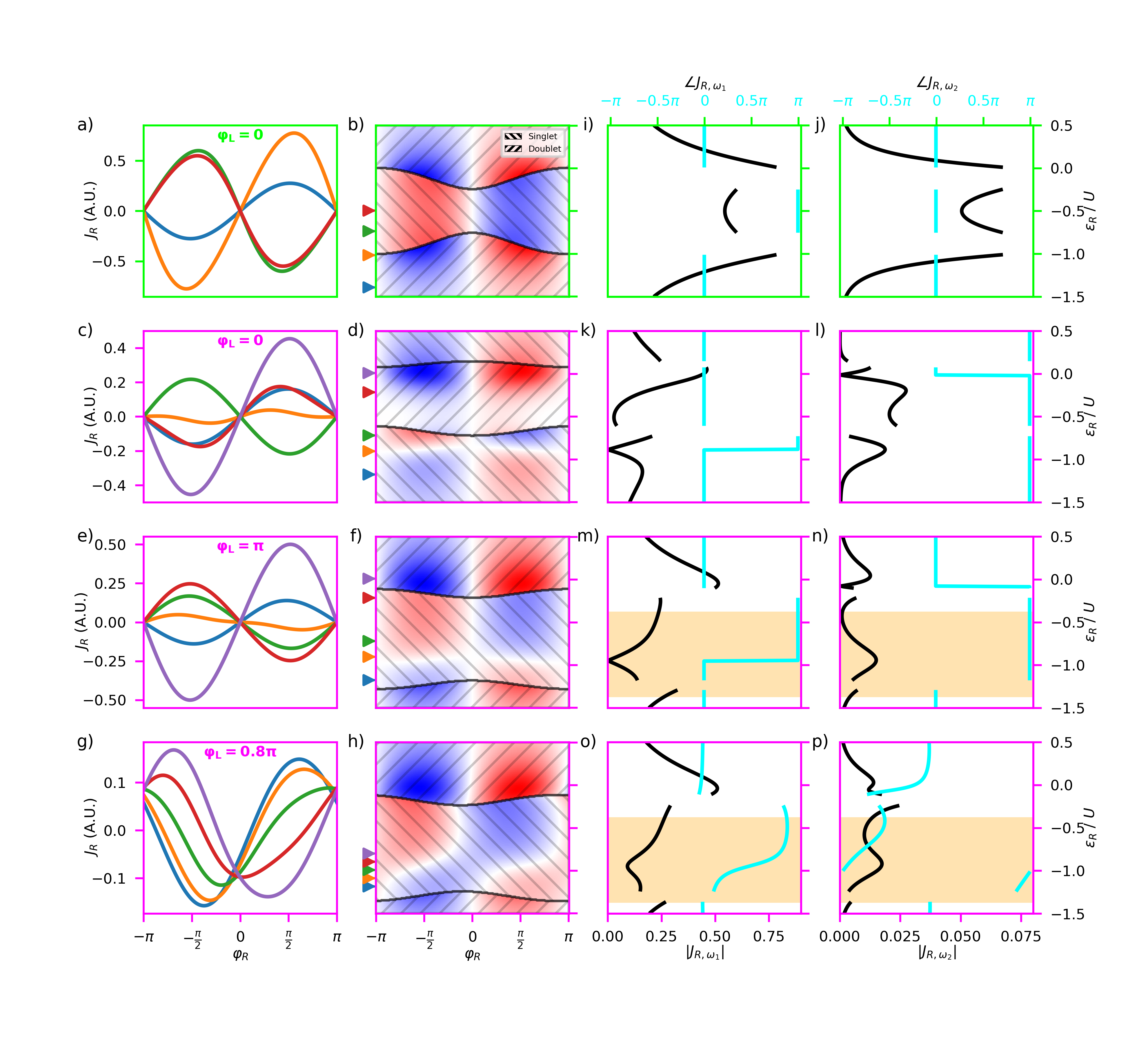}
	\caption[Caption]{
		Reproduction of Fig.~\ref{fig:cpr}, with the first and second harmonic parts of the CPR highlighted.
		The third column shows the first, the fourth column the second harmonic signal,
		with the black solid lines denoting the amplitude (bottom scale), the cyan lines the phase (top scale) of the signal.
		The lines are not continuous, where changing $\varphi_R$ induces a GS change, such as around $\varepsilon_R=0$ and $-1$ in the first row.
	}
	\label{figs:2w1}
\end{figure*}

\begin{figure*}[]
	\includegraphics{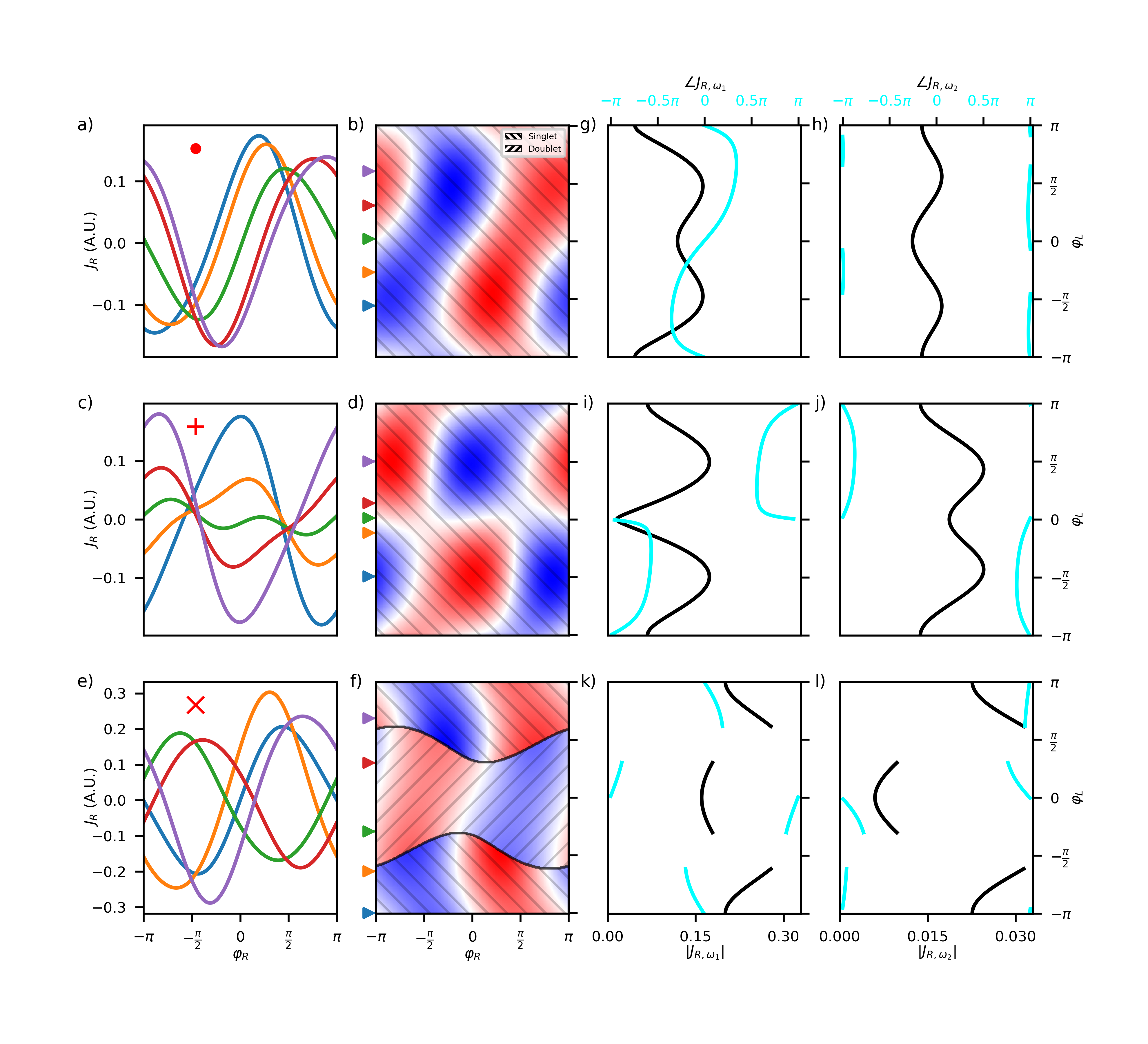}
	\caption[Caption]{
		Reproduction of Fig.~\ref{fig:nljf}, with the first and second harmonic parts of the CPR highlighted.
		The third column shows the first, the fourth column the second harmonic signal,
		with the black solid lines denoting the amplitude (bottom scale), the cyan lines the phase (top scale) of the signal.
		The lines are not continuous, where changing $\varphi_R$ induces a GS change, such as around $\varepsilon_R=0$ and $-1$ in the first row.
	}
	\label{figs:2w2}
\end{figure*}

In this section we show how the first and second harmonic component of $J_R$ are tuned separately.
This helps us gain a deeper insight into the $\pi$-periodic CPRs, as well as the $\varphi_0$ phase-shifts shown in the main text.
We are only concerned with the first two Fourier components, as the amplitude of higher harmonics is negligible.
This means that we can write the supercurrent as
$J_R\simeq J_{R,\omega_1}+J_{R,\omega_2}=\left|J_{R,\omega_1}\right|\sin\left(\varphi_R+\angle J_{R,\omega_1}\right)+\left|J_{R,\omega_2}\right|\sin\left(2\varphi_R+\angle J_{R,\omega_2}\right)$

As mentioned earlier, protected qubits based on systems with $\cos 2\varphi$ CPRs have been proposed\cite{schrade_protected_2022}.
This protection requires that $J_{\omega_1}=0$ and $J_{\omega_2}\neq0$.
However, if the suppression of the first harmonic signal is linear in a given parameter, $J_{\omega_1}(\alpha)\simeq \alpha$,
it will be sensitive to the noise of that parameter, and the protection is lost.
Ideally in a protected state the first harmonic component is suppressed at least quadratically in all parameters, $J_{\omega_1}(\alpha)\simeq \alpha^2$.
Thus we will look for such points in the parameter space.

Fig.~\ref{figs:2w1} is a reproduction of Fig.~\ref{fig:cpr}, with the addition of panels i-p.
The color of the axes corresponds to the value of $\varepsilon_L$ shown in Fig.~\ref{fig:eemap}, the labels in the first column show the value of $\varepsilon_L$.
The first column shows selected CPR curves, the supercurrent $J_R$ (left scale) as a function of the local phase $\varphi_R$ (bottom scale).
The second column shows the supercurrent $J_R$ as the function of the local phase $\varphi_R$ (bottom scale) and local on-site energy $\varepsilon_R$ (right scale).
The third column corresponds to the first harmonic part of the CPR,
the solid black line shows the amplitude (bottom scale) the cyan line the phase (top scale).
The fourth column is similar to the third, but shows the amplitude (black, bottom scale) and phase (cyan, top scale) of the second harmonic component of the signal.

\subsection{Off-resonance case}

The simplest case is that of the $0-\pi$ phase transition, when the nonlocal dot is off-resonance, as shown in the first row.
Starting at the bottom of Fig.~\ref{figs:2w1}b ($\varepsilon_R < -1 \mathrm{U}$, blue and orange curve of Fig.~\ref{figs:2w1}a),
we see that the phase of the first harmonic signal is 0.
Above $\varepsilon_R\simeq-1 \mathrm{U}$ the CPR, $J_R(\varphi_R)$, curve shows jumps as singlet-doublet phase transitions are triggered when sweeping $\varphi_R$,
so the Fourier decomposition of the CPR is not directly usable.
This is why the curves of Fig.~\ref{figs:2w1}i and j are not continuous for values of $\varepsilon_R$,
where changes in $\varphi_R$ can trigger singlet-doublet transitions.
Above $\varepsilon_R\simeq-0.75 \mathrm{U}$ (green curve and arrow on Fig.~\ref{figs:2w1}a and b) we see no jumps, but the CPRs are shifted by $\pi$,
as shown by the jump in the red curve of Fig.~\ref{figs:2w1}i.
Around $\varphi_R\simeq0$ we see a similar $0-\pi$ phase transition.
The Fourier analysis of the first harmonic signal shows that the phase-shift if indeed $\pi$, as described in the main text.
This shows that decomposing the CPR signal is a good tool to determine the exact phase shift.

We also note that while there is a non-negligible second harmonic signal even in the off-resonant case,
its amplitude is always smaller than that of the first harmonic part,
thus it only leads to the skewing of the signal.

\subsection{Hybridization}

The second and third row of Fig.~\ref{figs:2w1} show the nonlocal QD on resonance, and the nonlocal phase is $0$ of $\pi$.
In section \ref{sss:h1} and \ref{sss:h2} we describe how a $0-\pi$ phase shift occurs, without a singlet-doublet transition.
Fig.~\ref{figs:2w1}k and m show that the first harmonic singlet indeed has a $\pi$ phase jump,
when the amplitude goes to 0.
Examining the same locations on Fig.~\ref{figs:2w1}l and n,
we see that the second harmonic signal has a local maxima close to where the first harmonic goes to zero.
This explains why the orange curves of Fig.~\ref{figs:2w1}c and e seem to be $\pi$-periodic.
The amplitude of the second harmonic signal is higher in the case of the red curve of Fig.~\ref{figs:2w1}d,
but since the first harmonic signal is not zero, it only manifests as the skewness of the CPR.
The purple curve of the same panel has an even higher first harmonic component,
while the second harmonic is zero, leading to a pure sinusoidal signal.

Here we see that $J_{R,\omega_1}=0$ and $J_{R,\omega_2}\neq 0$ (e.g. close to $\varepsilon_R=-1\mathrm{U}$ on Fig.~\ref{figs:2w1}m and n),
however the first harmonic signal depends linearly on $\varepsilon_R$ close to its minimum,
thus making the system susceptible to gate noise, in contrast to the protection requirement of the $\cos 2\varphi$ qubit.

\subsection{$\varphi_0$ phase shift}

The fourth row of Fig.~\ref{figs:2w1} details the case of the $\varphi_0$ phase shift tuned by $\varepsilon_R$, when $\varphi_L=0.8\pi$.
Fig.~\ref{figs:2w1}o shows how the exact value of the $\varphi_0$ phase shift of the CPR depends on $\varepsilon_R$.
It is also noteworthy, that unlike in the case of $\varphi_L=0,\pi$, the amplitude of the first harmonic component of the CPR never goes to 0.
Even though the amplitude of the second harmonic signal still has a local maximum at the minimum of the first harmonic signal,
since the later does not go to 0, it only manifests as the skewness for the CPR curves.

\subsection{Non-local phase tuning}

For completeness we also reproduce Fig.~\ref{fig:nljf}, extended with the amplitude and phase of the first and second harmonic signals,
on Fig.~\ref{figs:2w2} to help us further explore the effects of phase-tuning.
Here we also see that when the first harmonic signal disappears (see Fig.~\ref{figs:2w2}i),
it does so linearly in $\varphi_L$, thus making the system susceptible to phase noise also.

\bibliography{main,extras}

\end{document}